\documentclass{article}
\usepackage{ijcai26}

\usepackage{times}
\usepackage{soul}
\usepackage{url}
\usepackage[hidelinks]{hyperref}
\usepackage[utf8]{inputenc}

\usepackage[small]{caption}
\usepackage{graphicx}
\usepackage{amsmath}
\usepackage{amsthm}
\usepackage{booktabs}
\usepackage{algorithm}
\usepackage{algorithmic}
\usepackage[switch]{lineno}
\usepackage{xcolor}
\usepackage{xspace}

\usepackage{float}
\usepackage{stfloats}
\usepackage[section]{placeins}
\usepackage{caption}
\usepackage{enumitem}

\newcommand{\facet}{FACET\xspace}

\title{FACET: Multi-Agent AI Supporting Teachers in Scaling Differentiated Learning for Diverse Students}

\author{
Jana Gonnermann-Müller$^{1,4}$\and
Jennifer Haase$^{3,4}$\and
Nicolas Leins$^1$\and
Moritz Igel$^1$\and
Konstantin Fackeldey$^{1,2}$\And
Sebastian Pokutta$^{1,2}$\\
\affiliations
$^1$Zuse Institute Berlin, Berlin, Germany\\
$^2$Technical University Berlin, Berlin, Germany\\
$^3$Humboldt University, Berlin, Germany\\
$^4$Weizenbaum Institute, Berlin, Germany\\
\emails
\{gonnermann-mueller, leins, igel, fackeldey, pokutta\}@zib.de,
haase.hu-berlin.de
}

\begin{document}

\maketitle

\begin{abstract}
Classrooms are becoming increasingly heterogeneous, comprising learners with diverse performance and motivation levels, language proficiencies, and learning differences such as dyslexia and ADHD. While teachers recognize the need for differentiated instruction, growing workloads create substantial barriers, making differentiated instruction an ideal that is often unrealized in practice. Current AI educational tools, which promise differentiated materials, are predominantly student-facing and performance-centric, ignoring other aspects that shape learning outcomes. 
We introduce \facet, a teacher-facing multi-agent framework designed to address these gaps by supporting differentiation that accounts for motivation, performance, and learning differences. Developed with educational stakeholders from the outset, the framework coordinates four specialized agents, including learner simulation, diagnostic assessment, material generation, and evaluation within a teacher-in-the-loop design.
School principals ($N = 30$) shaped system requirements through participatory workshops, while in-service K–12 teachers ($N = 70$) evaluated material quality. Mixed-methods evaluation demonstrates strong perceived value for inclusive differentiation. Practitioners emphasized both the urgent need arising from classroom heterogeneity and the importance of maintaining pedagogical autonomy as a prerequisite for adoption. We discuss implications for future school deployment and outline partnerships for longitudinal classroom implementation.
\end{abstract}

\section{Introduction}

Today's classrooms are increasingly diverse, with students varying widely in prior knowledge, motivation, language proficiency, and learning differences. This diversity includes a significant number of learners with neurodiversity: an estimated 5\% to 17\% of schoolchildren have dyslexia \cite{habibChapter23Dyslexia2013} and 11.4\% of U.S. children aged 3–17 years received an attention-deficit/hyperactivity disorder (ADHD) diagnosis in 2022 \cite{danielson_adhd_2024}. Addressing such heterogeneity requires differentiated instruction, for example, adapting content and support to individual learners. Yet differentiation significantly increases instructional complexity and teacher workload, creating a persistent gap between pedagogical ideals and what is feasible in everyday classroom practice.
Artificial intelligence (AI) has emerged as a potential solution to this challenge. AI-driven educational platforms, such as Khan Academy's Khanmigo in the United States \cite{khanacademyKhanAcademy2025} or Squirrel AI in China \cite{squirrelaiSquirrelAI2025} increasingly offer individualized learning paths. Yet these systems are predominantly student-facing, designed for direct learner interaction \cite{zhang_simulating_2025}, rather than teacher support. Moreover, their adaptation is typically driven by performance metrics alone, leaving them largely insensitive to the motivational barriers, affective responses, and processing differences that characterize diverse learners.

Educational research demonstrates that effective differentiation must extend beyond cognitive performance to include motivational factors and learning differences, such as reading and writing difficulties or ADHD \cite{bernackiSystematicReviewResearch2021,gayeWorkingMemoryMath2024a}. Evidence supports this approach: Stepwise cognitive strategies improve mathematics outcomes for students with ADHD \cite{iseman_cognitive_2011}, whose working memory deficits require specially designed materials \cite{gayeWorkingMemoryMath2024a}. However, implementing such comprehensive differentiation in everyday practice remains prohibitively time-intensive for most teachers.
To manage increasing workloads, some teachers turn to large language models (LLM) (e.g., ChatGPT) for generating teaching materials \cite{schummel_specifying_2025}. However, using single LLMs presents several limitations: First, they exhibit instability in learner simulation and high prompt sensitivity \cite{mannekoteCanLLMsReliably2025,ilkou_dyslexia_2025}. Second, they often fail to align with specific curricula, learning goals, or classroom contexts. Third, chatbots lack the scalability required for complex workflows, which becomes problematic when accommodating several learning groups and new requirements during the semester.
Multi-agent architectures offer a promising alternative to single LLM approaches \cite{wang_llm-powered_2025,haase_beyond_2025}. They enable stable, reusable learner representations through specialized agents with distinct pedagogical roles. Teachers can use the same student personas throughout the school year, reducing prompting effort. Additionally, agents can be equipped with structured knowledge, such as curriculum-aligned tasks and learning goals, didactic approaches, and guidelines for designing accessible teaching materials for students with dyslexia, ADHD, and other learning requirements. 

\begin{figure*}[th]
    \centering
    \includegraphics[width=1\linewidth]{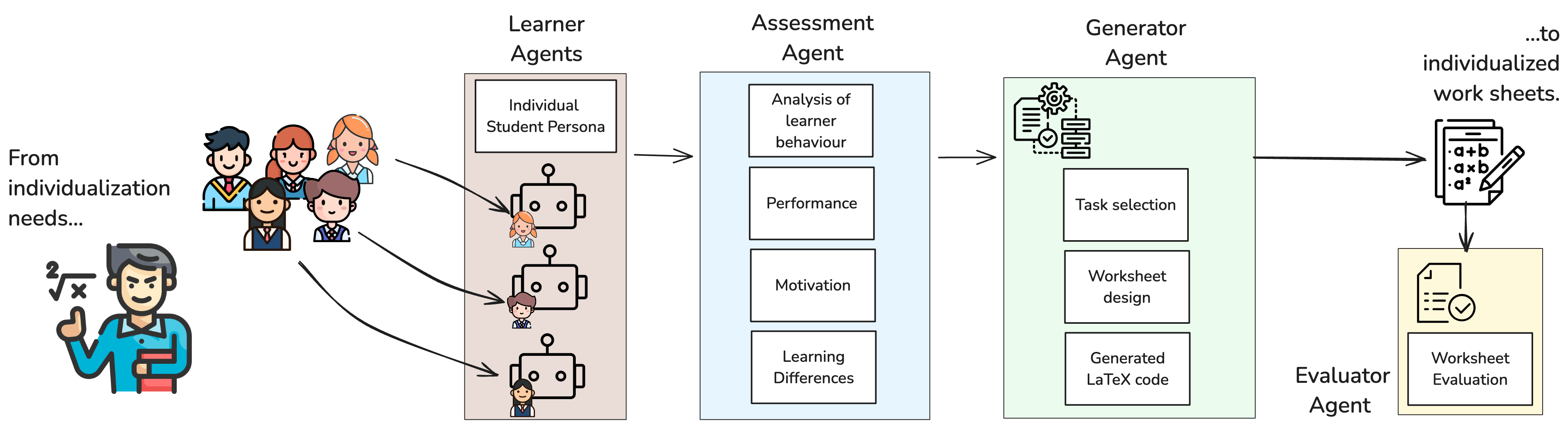}
    \caption{Core Principle for individualized classroom material using a multi-agent framework}
    \label{fig:banner}
\end{figure*}

Building on these advances, we present \facet, a teacher-facing multi-agent system designed to operationalize differentiation beyond performance while preserving educator authority (see Figure~\ref{fig:banner}). Teachers define learner profiles based on diverse characteristics, including motivation, learning differences, and prior knowledge, modify uploaded materials or generate new content, and review all outputs before classroom use. This approach keeps teachers in control, ensuring that AI-generated materials serve their pedagogical goals and fit their classroom realities. Therefore, \facet was developed through close collaboration with domain experts; from the beginning, mathematics teachers informed curriculum alignment and validated the didactic structure of the generated worksheets. Teachers with expertise in dyslexia reviewed accessibility adaptations for dyslexia-related profiles. This close partnership with practitioners extends throughout our research, from the formative workshops and teacher surveys presented in this paper to an ongoing longitudinal deployment across schools, ensuring that \facet remains grounded in authentic classroom needs.

\paragraph{Our key contributions are:}

\begin{enumerate}[nosep,leftmargin=*]
    \item \facet, a modular four-agent architecture that separates learner simulation, diagnostic reasoning, material generation, and evaluation. 
    \item An operationalization of learner differentiation beyond performance through teacher-specifiable profiles, including cognitive and motivational aspects, as well as neurodiversity.
    \item Mixed-methods evaluation with school principals ($N=30$) and in-service teachers ($N=70$) assessing workflow integration and material quality.
\end{enumerate}

\section{Related Work}
Relevant work spans three areas: differentiated and inclusive instruction, multi-agent systems in education, and teacher-facing AI systems.

\subsection{Differentiated and Inclusive Instruction}
Research consistently shows that effective differentiation must address cognitive proficiency, motivation, emotional regulation, and needs related to neurodiversity to improve learning outcomes and motivation \cite{bernackiSystematicReviewResearch2021,farianiSystematicLiteratureReview2023}. Self-Determination Theory highlights the importance of supporting autonomy, competence, and relatedness for sustained engagement \cite{Deci2000-pj}, which are needs that performance-based metrics fail to capture. 
Neurodiversity refers to the natural variation in human brain function, often used in the context of autism spectrum disorder, ADHD, and learning disabilities. Two common examples of neurodivergence in school contexts are ADHD and dyslexia. ADHD is characterized by developmentally inappropriate levels of inattention, hyperactivity, and impulsivity that persist over time \cite{world_health_organisation_who_international_2025}. Dyslexia is a specific learning disability characterized by difficulties with accurate and fluent word recognition, as well as poor spelling and decoding, typically stemming from a deficit in the phonological component of language, with potential secondary consequences for reading comprehension and broader academic development \cite{carroll_toward_2025}. 

For learners with dyslexia or ADHD, learning and engagement depend heavily on instructional clarity, structure, and motivational support. Reading difficulties, language-processing challenges, and attention deficits can increase cognitive load, leading to reduced learning outcomes and disengagement.
These learner needs translate into concrete requirements for differentiated instructional design. Effective differentiation requires deliberate design choices, including task sequencing, scaffolding, and instructional structure that go beyond content adaptation \cite{ilkou_dyslexia_2025}. 
For learners with dyslexia, research- and practice-oriented guidelines emphasize the importance of readable, well-structured worksheets that reduce extraneous cognitive load \cite{relloGoodFontsDyslexia2013}. 
Scaffolding strategies, such as worked examples, stepwise guidance, and partially completed representations, support learners’ progression toward independent problem solving. Equally important are motivational cues that sustain engagement and reinforce self-efficacy. Bloom’s revised taxonomy provides a structured framework for guiding learners from basic recall to higher-order thinking \cite{andersonTaxonomyLearningTeaching2001}, enabling systematic task differentiation that balances challenge and support. Without support, implementing differentiation in everyday classrooms that incorporates more aspects than performance remains challenging for teachers due to time constraints and the limited availability of curriculum-aligned, learner-sensitive materials.

\subsection{LLM and Multi-Agent Systems for Differentiated Education}
Recent advances in LLMs demonstrate that they can simulate stable personality traits, motivational dispositions, and affective responses with high internal consistency \cite{aher_using_2023,wu_embracing_2025}. These capabilities enable the construction of reusable, parameterized learner profiles \cite{wang_evaluating_2025,huang_designing_2026} that overcome the scalability limitations of earlier AI-based tutoring systems and single LLM approaches for teachers. Building on this foundation, emerging research explores LLM-based digital twins and multi-agent classroom simulations in which multiple agents represent heterogeneous learners, instructional roles, or pedagogical functions \cite{zhang_simulating_2025,wu_embracing_2025}. Multi-agent architectures enable the separation of subtasks across specialized agents (e.g., learner simulation, pedagogical reasoning, assessment, adaptation), allowing intermediate reasoning steps to be inspected and corrected. Recent work demonstrates examples of student-facing multi-agent systems, such as the coordination of specialized agents for educational assessment \cite{hou_llm-enhanced_2025}; SimClass, a simulated classroom \cite{zhang_simulating_2025} and GenMentor, which coordinates multiple agents for goal mapping, skill gap identification, and adaptive content generation \cite{wang_llm-powered_2025}. Yet, similar to AI-driven educational platforms, existing multi-agent systems remain predominantly student-facing, designed for direct learner interaction rather than teacher support.

\subsection{Teacher-Facing AI Systems}
Multi-agent teacher-facing AI systems have progressed more slowly than student-facing applications. Existing tools typically provide static instructional resources or limited customization, often requiring substantial manual effort. Recent work on LLM-based lesson planning suggests that AI can meaningfully reduce teachers' planning workload \cite{huExploringPotentialLLM2025}, yet these systems rarely incorporate explicit learner modeling and differentiation that includes cognitive and motivational aspects, as well as neurodiversity.

Notable exceptions are emerging but remain domain-specific. StoryLab \cite{li_storylab_2025} demonstrates teacher-in-the-loop personalization for literacy, and ARISE \cite{gajewska_leveraging_2025} employs multi-agent simulation for teacher training. Neither supports differentiated material generation across diverse learner profiles.

\subsection{Synthesis}
Existing work demonstrates that effective differentiation requires considering aspects beyond performance alone, including motivational factors and learning differences such as dyslexia and ADHD. Multi-agent architectures provide the foundational infrastructure to enable stable, reusable learner representations, the integration of knowledge bases ensuring curriculum alignment and didactic guidelines, and longitudinal learning as the system accumulates context from materials uploaded throughout the school year. However, teacher-facing systems have received limited attention in research. We translate the capabilities of agentic AI into a teacher-facing tool that produces curriculum-aligned, accessibility-compliant materials while preserving teacher authority. In close collaboration with schools, we developed \facet to address these gaps. 

\section{\facet: A Teacher-Facing Multi-Agent Framework} 

\begin{figure*}[t]
    \centering
    \includegraphics[height=0.3\textheight]{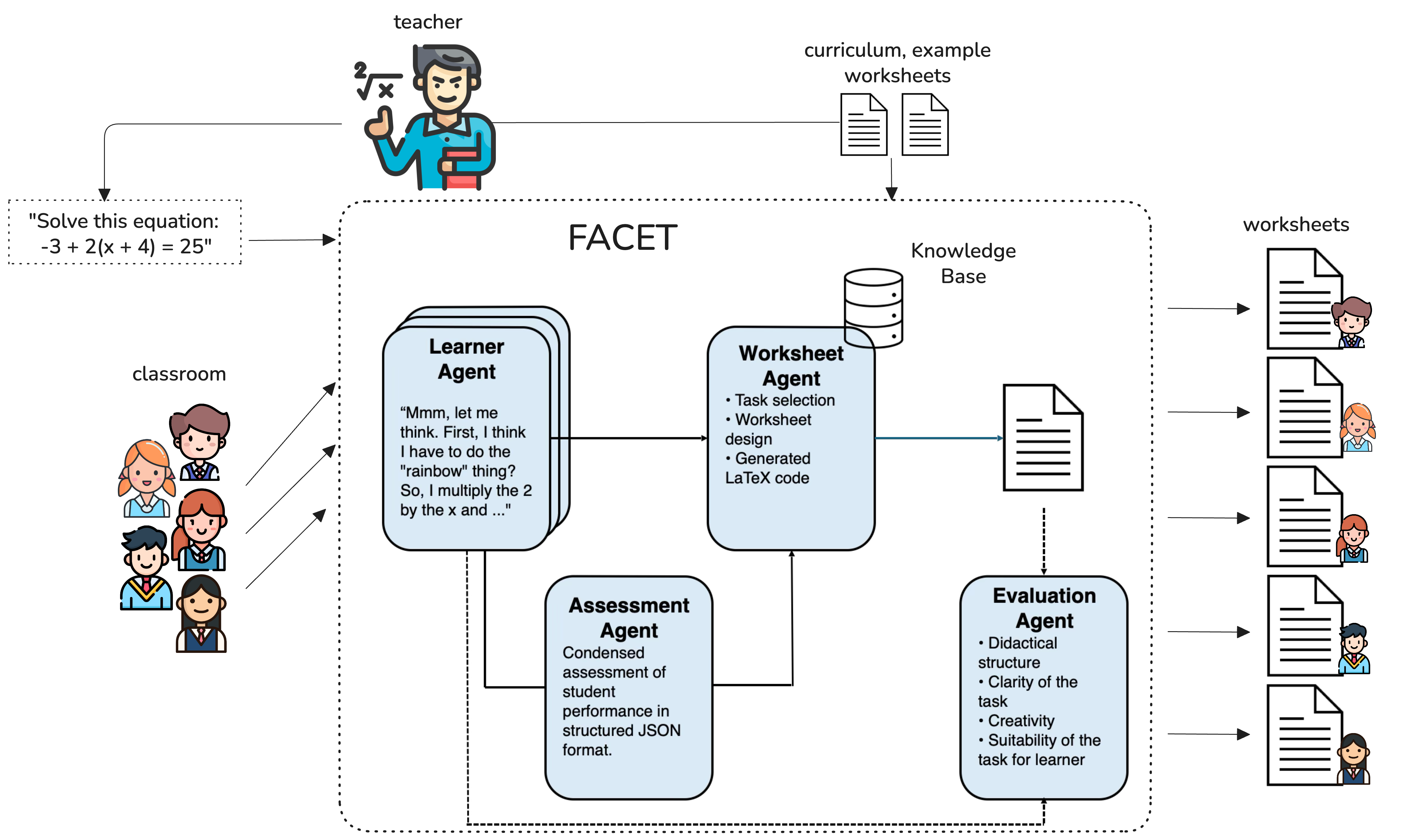}
    \caption{Core workflow: Teacher-defined learner profiles drive multi-agent generation of differentiated materials with human review}
    \label{fig:workflow}
\end{figure*}

\subsection{Design Rationale}
\facet adopts a multi-agent framework that addresses the limitations of single LLM approaches through three foundational capabilities. 
First, the system maintains stable learner representations through dedicated persona agents that consistently preserve learner characteristics across sessions, reducing the prompt sensitivity and output variability observed in single LLM approaches. This reduces the prompting burden on educators who are typically not prompt engineering experts. 
Second, the architecture enables the integration of structured knowledge bases, ensuring curriculum alignment and adherence to didactic guidelines in generated worksheets. Therefore, the generator agent is equipped with foundational knowledge, such as the curriculum and evidence-based didactic theories that are embedded in the system and can be extended by teachers through uploading their own teaching materials.  Third, the system supports longitudinal context accumulation, building understanding from materials that teachers upload throughout the school year. As the school year progresses, the system accumulates contextual knowledge about completed content and student progress, enabling increasingly personalized and curriculum-aligned differentiation while maintaining pedagogical rigor. 

\subsection{Agent Specifications}
The system architecture rests on a Python FastAPI backend coupled with a React frontend, ensuring a responsive teacher-in-the-loop experience. To guarantee data privacy and control over the generation process, we employ \textit{GPT-OSS 120B}, an open-source LLM from OpenAI hosted on local infrastructure via Ollama. 

\facet utilizes custom logic to orchestrate agent interactions, enforcing a strict pedagogical workflow, as illustrated in Figure~\ref{fig:workflow}. Communication between agents follows a predefined protocol to ensure stability: The learner agent generates natural language thought traces (thought protocol) of the student working on a task. These traces are passed to the assessment agent, which processes them into a strict JSON diagnostic format defined via a JSON schema in the system prompt. The generator agent enables the creation of the worksheet by receiving both the learner's thought protocol and the structured diagnostic JSON. In contrast, the evaluator agent functions as an external reviewer; it receives only the generated worksheet and the original learner persona prompt to assess suitability, operating without access to the internal reasoning steps or intermediate diagnostics. Expert knowledge (curriculum standards, didactic guidelines) is integrated via context injection (system prompts), with plans to implement a Retrieval-Augmented Generation (RAG) pipeline to overcome context window limitations and allow for scalable knowledge management in future iterations. To balance creativity with reliability, we apply a temperature of 0.9 to the learner, assessment, and generator agents, while restricting the evaluator agent to 0.1 for consistent diagnostic reasoning. The full generation cycle (simulation, diagnosis, and material creation) for one worksheet completes in approximately 30 seconds.

\subsection{Learner Agents: Simulating Diverse Learner Profiles}
Teachers can create student descriptions that reflect the characteristics they commonly observe in their classrooms. Consistent with our framing of differentiation beyond performance, these profiles integrate diverse dimensions, such as prior knowledge levels, motivational aspects, reading and writing difficulties, attention-related needs, and other cognitive and affective factors. The learner agents simulate student behavior based on teacher-defined characteristics. At the beginning of each process, the learner agent attempts to solve an assigned task while generating a structured transcript that captures reasoning steps, misconceptions, and emotional responses. Tasks can be uploaded by teachers or selected from curriculum-aligned repositories. By leveraging LLMs' demonstrated capacity to simulate consistent personality traits and affective patterns \cite{aher_using_2023,li_quantifying_2024}, \facet approximates realistic learner behavior as a basis for instructional adaptation.

We emphasize that learner agents are not intended to serve as clinical or diagnostic models for dyslexia or ADHD. Rather, they function as controllable simulations that test instructional materials against specified constraints (e.g., reduced text density, shorter instructions, and higher structure). To mitigate stereotyping risks, profiles are defined in terms of observable, learning-relevant needs, such as decoding difficulty, working-memory load, or sustained attention demands, rather than diagnostic labels alone. 

\subsection{Assessment Agent: Diagnostic Reasoning}
The assessment agent analyzes simulated learner interactions with instructional materials, examining both cognitive processes and affective responses. Guided by a strict JSON schema provided in the system prompt to ensure structural consistency, it identifies indicators such as conceptual misunderstandings, low confidence, disengagement, or, conversely, fluent performance and high motivation. The resulting structured diagnostic summary provides an explicit foundation for subsequent generation and conditioning material adaptation based on identified needs.

\subsection{Generator Agent: Differentiated Worksheets}
The generator agent creates differentiated instructional materials grounded in the diagnostics produced by the preceding agents. It is equipped with structured knowledge bases that ensure both curriculum alignment and adherence to didactic principles. Teachers can specify high-level directives, such as instructional style, task selection strategies, and worksheet format, allowing \facet to align with individual pedagogical preferences while the agent contributes design decisions informed by established instructional frameworks embedded in the system or extended through teacher-uploaded materials. Critically, the generator incorporates accessibility guidelines identified in research as essential for neurodivergent learners, such as clear sans-serif fonts, sufficient line spacing, strong text–background contrast, single-column layouts, and explicit visual separation of tasks \cite{relloGoodFontsDyslexia2013}. Scaffolding strategies, worked examples, stepwise guidance, and partially completed representations are integrated based on Bloom's revised taxonomy \cite{andersonTaxonomyLearningTeaching2001}, a framework for progressing from foundational to complex cognitive tasks to systematically balance challenge and support.
The decision to generate structured \LaTeX{} code serves a dual purpose: it provides a robust, machine-readable format that ensures precise layout control for accessibility compliance (e.g., specific fonts, spacing), while simultaneously enabling flexible export options. The system directly renders professional-grade PDFs or converts the source to \texttt{.docx} format, allowing teachers unfamiliar with \LaTeX{} to easily edit materials in standard word processors. Figure~\ref{fig:worksheet_example} in the Appendix shows an example worksheet.

\subsection{Evaluator Agent: Pedagogical Quality Assurance}
The evaluator agent provides structured feedback on generated materials along four pedagogically relevant dimensions aligned with our evaluation framework: Didactic structure, clarity, creativity, and format and design. This evaluation serves a dual purpose. It informs the iterative refinement of the system while providing transparent and easy-to-understand feedback to teachers.
 
\subsection{Teacher-In-The-Loop Interface}
Consistent with the teacher-in-the-loop principle, educators retain full control throughout the workflow. Teachers interact with \facet at three points: (1) profile specification via prompt templates or self-written prompts, (2) upload of materials or task selection from curriculum-aligned repositories, and (3) output review with accept/modify/reject controls. Teachers can decide to export materials as PDF, Word, or \LaTeX{}. This design reflects our core position: \facet augments teachers by scaling the production of differentiated materials, freeing cognitive and temporal resources for responsive instruction and meaningful student interaction, rather than replacing the contextual knowledge and professional judgment that only educators possess. Figure~\ref{fig:workflow} depicts the workflow. 

\section{Evaluation Methods} 
We evaluated \facet through two complementary stakeholder engagements. First, we conducted exploratory workshops with German school principals who will prospectively deploy the system in their schools, ensuring alignment with authentic institutional needs and workflow constraints. Second, to assess generalizability and broaden quality assessment beyond a single educational context, we surveyed practicing teachers from countries with high-performing education systems: the United Kingdom, the United States, Canada, and Australia. This dual approach addresses two research questions: 

\begin{description}[nosep]
    \item[RQ1:] To what extent does \facet address teachers' practical needs and integrate with existing workflows?
    \item[RQ2:] To what extent can \facet generate pedagogically suitable materials for diverse learner profiles?
\end{description}

\subsection{Workshop Study (RQ1)}
Through two one-hour workshops with school principals ($N=30$, 15 per session) from German schools, we examined practical relevance, usability, and classroom integration. After a tool demonstration, participants engaged in structured discussions on system structure and usage, interaction possibilities and degrees of freedom, and intended output. Each session was led by one facilitator, with two researchers documenting participant responses.

\subsection{Survey Study (RQ2)}
Building on the workshop findings, we conducted a quantitative online survey with K–12 in-service teachers recruited via Prolific. We derived tasks from the grade 7 mathematics curriculum (area calculation) to ensure ecological validity. Teachers were presented with five example worksheets generated by \facet, each corresponding to a distinct learner profile with varying motivations, performance, or dyslexia: 

\begin{itemize}
    \item Profiles that represent varying combinations of mathematical proficiency and motivational orientation.
    \item Profiles that represent learners with dyslexia-related processing needs, specified in terms of observable learning-relevant characteristics.
\end{itemize}

The selection of student characteristics was informed by discussions with participating teachers during the workshops and reflects learner characteristics commonly observed in heterogeneous classrooms. All profiles used gender-neutral language to reduce potential confounding effects. 

Participants were asked to review all five worksheets and evaluate them on a six-point Likert scale (1 = very low quality, 6 = very high quality). Each worksheet was rated along four pedagogically relevant dimensions: (1) didactic structure: How well is the worksheet structured from an educational perspective (e.g., support, sequence of tasks), (2) clarity of task formulation: How comprehensible are the tasks and the language?, (3) format and design: How appealing, clear, and reader-friendly is the worksheet? Does the layout support motivation and working with the material?, and (4) creativity: How original, motivating, and varied is the design of the tasks compared to typical teaching materials. For worksheet 5, additional items addressed font readability, language simplification, and spacing/layout. 

Survey data was analyzed descriptively by computing means and standard deviations for each evaluation dimension across worksheets and learner profiles. 

\section{Results} 
This section presents findings from our mixed-methods evaluation addressing the two research questions. 

\subsection{Workshop Findings (RQ1)} 
The thematic analysis of workshop sessions ($N=30$ school principals) identified three recurring themes regarding practical integration: practical relevance, usability and teacher-oriented design, and classroom integration and use cases. Table~\ref{tab:workshop_feedback} summarizes the findings. Participants emphasize the strong need for AI-based differentiation support, noting that inclusive classrooms increasingly require addressing diverse needs (ADHD, learning difficulties, social-emotional challenges) that exceed individual teacher capacity. School principals highlighted the pedagogical value of supporting both downward and upward differentiation, the importance of curriculum alignment and accessibility (e.g., for learners with dyslexia), and the necessity of reducing teacher workload amid conditions of teacher shortages and diverse learner needs. Accessible formatting for dyslexia, such as using clear layouts and reduced text density, was identified as critical. The teacher-in-the-loop design was viewed positively, as it preserves professional judgment, a prerequisite for trust and adoption.

\begin{table*}[t]
\centering
\caption{Key themes from exploratory workshops with school principals ($N = 30$)}
\label{tab:workshop_feedback}
\small
\begin{tabular*}{\textwidth}{@{\extracolsep{\fill}} p{0.2\textwidth} p{0.75\textwidth}}
\toprule

\multicolumn{2}{l}{\textit{Practical Relevance}} \\
\midrule

\textbf{Differentiation Support} &
We need both scaffolding for struggling learners and enrichment for advanced students. But realistically, when time runs out, we end up just giving extra worksheets or moving students to the next topic, what we know that's not ideal, but what else can we do? \\

\textbf{Teacher Workload} &
I have 34 students with completely different needs—ADHD, autism, learning difficulties, emotional challenges. With the teacher shortage, we're stretched thin. Differentiation isn't optional anymore, but it's impossible to sustain without help. \\

\textbf{Curriculum Alignment} &
If the AI generates tasks that don't match our curriculum or miss the right difficulty level, I won't use it. The materials have to be something I could actually hand out in class tomorrow. \\

\textbf{Accessibility} &
For my students with dyslexia, formatting matters enormously. Clear layouts, less text on the page, explicit structure. These aren't nice-to-haves, they're essential for those kids to even access the content. \\

\midrule
\multicolumn{2}{l}{\textit{Usability and Teacher-Oriented Design}} \\
\midrule

\textbf{Teacher-in-the-Loop Control} &
I need to stay in control. If the system generates something and I can't review it, change it, or throw it out—I won't trust it. I'm the one who knows my students. \\

\textbf{Customization and Consistency} &
My worksheets have a certain look; e.g. my symbols, my structure. Students know what to expect. Any new tool has to fit into that, not force me to start from scratch. \\

\textbf{Configurable Learners} &
I don't want to write prompts from scratch every time. Give me templates I can adjust—pre-set profiles for different performance levels that I can tweak based on what I know about my class. \\

\midrule
\multicolumn{2}{l}{\textit{Classroom Integration and Use Cases}} \\
\midrule

\textbf{Instructional Orchestration} &
I need materials I can use flexibly, for the whole class, for group work, for individual practice. And please, automatic answer keys. I don't have time to create those separately. \\

\textbf{Transparency of AI Reasoning} &
I like seeing the different learner personas. This makes me think about why a task is designed a certain way. But we have to be careful not to reduce students to stereotypes. \\

\textbf{Extension to Subjects} &
This would be incredibly useful in other subjects too, e.g., languages, history. And we need profiles for ADHD students, not just for dyslexia. The more flexible, the better. \\

\bottomrule
\end{tabular*}

\vspace{0.3em}
\raggedright
\footnotesize{\textit{Note. Representative quotes synthesized from two exploratory workshops with German school principals ($N = 30$).}}
\end{table*}

\subsection{Survey: Material Quality Assessment (RQ2)}
Seventy K--12 teachers ($M_{\text{age}}=40.5$, $SD=9.5$; 73\% female) evaluated five example worksheets generated by \facet. The majority of participants resided in the United Kingdom (57.14\%) or the United States (34.29\%), with smaller proportions from Canada (7.14\%) and Australia (1.43\%). Regarding school type, approximately half of the participants taught at primary schools (48.57\%), while 42.86\% taught at secondary schools, and 77.14\% had experience teaching students aged 11–13 either currently or within the past two years, which reflects expertise in the age group for which the example worksheets were generated. Notably, participants demonstrated high openness toward AI integration, with 91.43\% expressing a willingness to use AI for lesson preparation and 70.0\% already incorporating AI into their lesson preparation practices. Table~\ref{tab:results} presents the descriptive statistics for the worksheet evaluation.

\begin{table}[h]
\centering
\caption{Descriptive Statistics for Worksheet Evaluation}
\label{tab:results}
\small
\begin{tabular}{lrrr}
\toprule
Criterion & $M$ & $SD$ & Mdn \\
\midrule
\multicolumn{4}{l}{\textit{WS 1: High Performance, low Motivation}} \\
\quad Didactic Structure & 4.50 & 1.02 & 5 \\
\quad Clarity of Task Formulation & 4.46 & 1.05 & 5 \\
\quad Format and Design & 3.79 & 1.36 & 4 \\
\quad Creativity & 3.66 & 1.27 & 4 \\
\addlinespace[0.3em]
\multicolumn{4}{l}{\textit{WS 2: Low Performance, low Motivation}} \\
\quad Didactic Structure & 4.41 & 1.06 & 4 \\
\quad Clarity of Task Formulation & 4.23 & 1.16 & 4 \\
\quad Format and Design & 3.63 & 1.35 & 4 \\
\quad Creativity & 3.50 & 1.41 & 4 \\
\addlinespace[0.3em]
\multicolumn{4}{l}{\textit{WS 3: High Performance, high Motivation}} \\
\quad Didactic Structure & 3.87 & 1.17 & 4 \\
\quad Clarity of Task Formulation & 3.57 & 1.20 & 4 \\
\quad Format and Design & 3.10 & 1.37 & 3 \\
\quad Creativity & 3.30 & 1.34 & 3 \\
\addlinespace[0.3em]
\multicolumn{4}{l}{\textit{WS 4: Low Performance, high Motivation}} \\
\quad Didactic Structure & 4.03 & 1.20 & 4 \\
\quad Clarity of Task Formulation & 3.80 & 1.19 & 4 \\
\quad Format and Design & 3.29 & 1.46 & 3.5 \\
\quad Creativity & 3.40 & 1.36 & 3 \\
\addlinespace[0.3em]
\multicolumn{4}{l}{\textit{WS 5: Dyslexia Profile}} \\
\quad Suitability for Reading Difficulties & 3.46 & 1.47 & 3.5 \\
\quad Font Readability & 3.30 & 1.61 & 3 \\
\quad Language Simplification & 3.94 & 1.40 & 4 \\
\quad Spacing and Layout & 4.07 & 1.39 & 4 \\
\bottomrule
\end{tabular}

\vspace{0.5em}
\raggedright
\footnotesize
\textit{Note.  $N = 70$. Items rated on a 6-point scale (1 = \textit{very low}, 6 = \textit{very high}).}
\end{table}

\paragraph{Core quality dimensions.}
Didactic structure and clarity received consistently favorable ratings, exceeding the scale midpoint across worksheets for profiles 1--4 (range: $3.5-4.50$). Worksheets targeting low-motivation learners (W1, W2) achieved the highest ratings for didactic structure ($M=4.41-4.50$) and clarity ($M=4.23-4.46$), indicating effective scaffolding and motivational support.

\paragraph{Enrichment materials.}
Worksheet 3 (high-performing, highly motivated) received lower ratings across all dimensions (structure: $M=3.87$; clarity: $M=3.57$; format: $M=3.10$; creativity: $M=3.30$). This pattern suggests that the framework currently generates scaffolded content more effectively than enrichment-oriented materials for advanced learners.

\paragraph{Accessibility adaptation.}
The dyslexia-adapted worksheet (W5) yielded mixed results. Spacing/layout ($M=4.07$) and language simplification ($M=3.94$) exceeded the midpoint, indicating partial alignment with accessibility guidelines. However, font readability ($M=3.30$) and overall suitability for reading difficulties ($M=3.46$) fell near the midpoint, revealing that typographical requirements for dyslexia can be improved. This finding identifies a specific improvement target for the generation pipeline.

\paragraph{Format and creativity.}
Format/design (range: 3.10--3.79) and creativity (range: 3.30--3.66) received moderate ratings near the midpoint. As noted in the survey instructions, teachers were asked to focus on pedagogical content rather than visual appearance, since the system allows teachers to upload their own example materials to customize the format and style of generated outputs. Nevertheless, these dimensions represent opportunities for enhancement.

Overall, worksheets received favorable ratings, with most criteria scoring around 4 out of 6 and some reaching 5 out of 6, particularly for didactic structure and clarity of task formulation. However, results revealed specific areas for improvement: worksheets for high-motivation profiles (both low and high performance) received lower ratings for creativity and format/design, suggesting a need for more engaging and varied materials for motivated learners. The dyslexia-adapted worksheet showed room for improvement in font readability and suitability for reading difficulties, indicating that typographical adaptations require further refinement to meet accessibility guidelines.

\section{Limitations and Future Work}

We identified four limitations informing our continued stakeholder collaboration:
Our evaluation focused on grade-7 mathematics, demonstrating feasibility without establishing cross-class and cross-subject generalizability. Furthermore, agents simulate behavior based on LLM trait consistency, which requires further validation with authentic student responses.

We address these limitations through continued stakeholder collaboration in longitudinal deployment:

\paragraph{School partnerships.} We have established collaborations with 10 German schools, primary, secondary, and inclusive schools, for deployment. Five teachers per school will integrate \facet into regular instruction over one semester, enabling the evaluation of workflow integration, teachers' trust in the system, and student feedback under authentic conditions.

\paragraph{Embedded validation mechanisms.} Within deployment, teachers will systematically compare simulated learner behavior against actual student responses, flagging implausible patterns and documenting divergences. This practitioner-driven validation grounds simulation refinement in classroom reality.

\paragraph{Student feedback.} The diverse school sample spans socioeconomic contexts, enabling an analysis of how different students perceive material quality. 

\section{Conclusion}
We present \facet, a teacher-facing multi-agent system that addresses a pressing societal challenge: enabling differentiated instruction at scale in heterogeneous classrooms where teachers face unsustainable workload demands. Our four-agent architecture, consisting of learner simulation, diagnostic reasoning, material generation, and evaluation, operationalizes differentiation that incorporates aspects beyond performance-based adaptation while preserving educator authority.
Our mixed-methods evaluation confirms that \facet meets genuine classroom needs. School principals validated that the framework addresses real constraints, emphasizing the importance of teacher-in-the-loop control, configurable profiles, and curriculum alignment. Teachers rated scaffolding-oriented materials as classroom-ready, with good didactic structure and clarity. However, enrichment materials for advanced learners and dyslexia-specific formatting require further refinement.
Developed through sustained collaboration with educational stakeholders, \facet demonstrates that AI can meaningfully support differentiated instruction when designed around teachers' needs. These findings, along with identified improvement targets, inform our ongoing longitudinal deployment across partner schools.

\section*{Ethical Statement}
FACET is a research framework that simulates learner profiles and generates differentiated materials to support teachers. Its use requires clear human oversight at all stages. All study data is anonymized/pseudonymized where possible, and participation requires informed consent that explains the simulated classroom setting, AI-generated content, and data collection; participation is voluntary and appropriately compensated. We acknowledge that persona mimicking could be misunderstood. To address this, personas are treated as analytic abstractions rather than representations of real individuals, all outputs are labeled as AI-generated, and teachers retain full control to edit, reject, or withhold materials. Nevertheless, our commitment to data privacy is reinforced by a fully local infrastructure. While the framework is accessible via the web to international users, all application servers and AI inference models are hosted on-site at the authors’ institute. To maintain full data sovereignty, the system does not utilize third-party cloud providers or proprietary LLM APIs; all data is stored and processed exclusively on our self-hosted institutional servers and remains within our local network. It is not yet empirically established how students will perceive AI-adapted worksheets in real classrooms, including whether differentiation could feel stigmatizing, isolating, or like “special treatment.” Therefore, student-facing use is conditional on rigorous evaluation that includes psychosocial outcomes such as perceived fairness, autonomy, belonging, and self-esteem, not only performance. We further recognize the risk that simulated student representations may be unfair, diminishing, or negatively biased, potentially reinforcing deficit views and shaping teacher expectations. To mitigate this, we avoid derogatory descriptors and protected-attribute profiling, and we implement bias-aware prompting, auditing, and teacher-in-the-loop review. Because LLM systems can hallucinate or produce incorrect information, FACET outputs must be treated as drafts requiring verification, especially in high-stakes contexts. Finally, AI agents cannot replace real teachers in providing emotional support, social learning, and normative guidance, and any deployment should be interdisciplinary, transparent, and continuously monitored for unintended harm.

\section*{Acknowledgments}
We thank the Zuse Institute Berlin (\url{https://www.zib.de}) for hosting the system and LLMs. Research reported in this paper was partially supported through the Research Campus Modal, funded by the German Federal Ministry of Research, Technology and Space (fund numbers 05M14ZAM,05M20ZBM) and by the German Federal Ministry of Research, Technology and Space, grant number 16DII133 (Weizenbaum-Institute). 

\newpage

\section*{Appendix}

\begin{figure}[h]
    \centering
    \includegraphics[width=0.85\columnwidth]{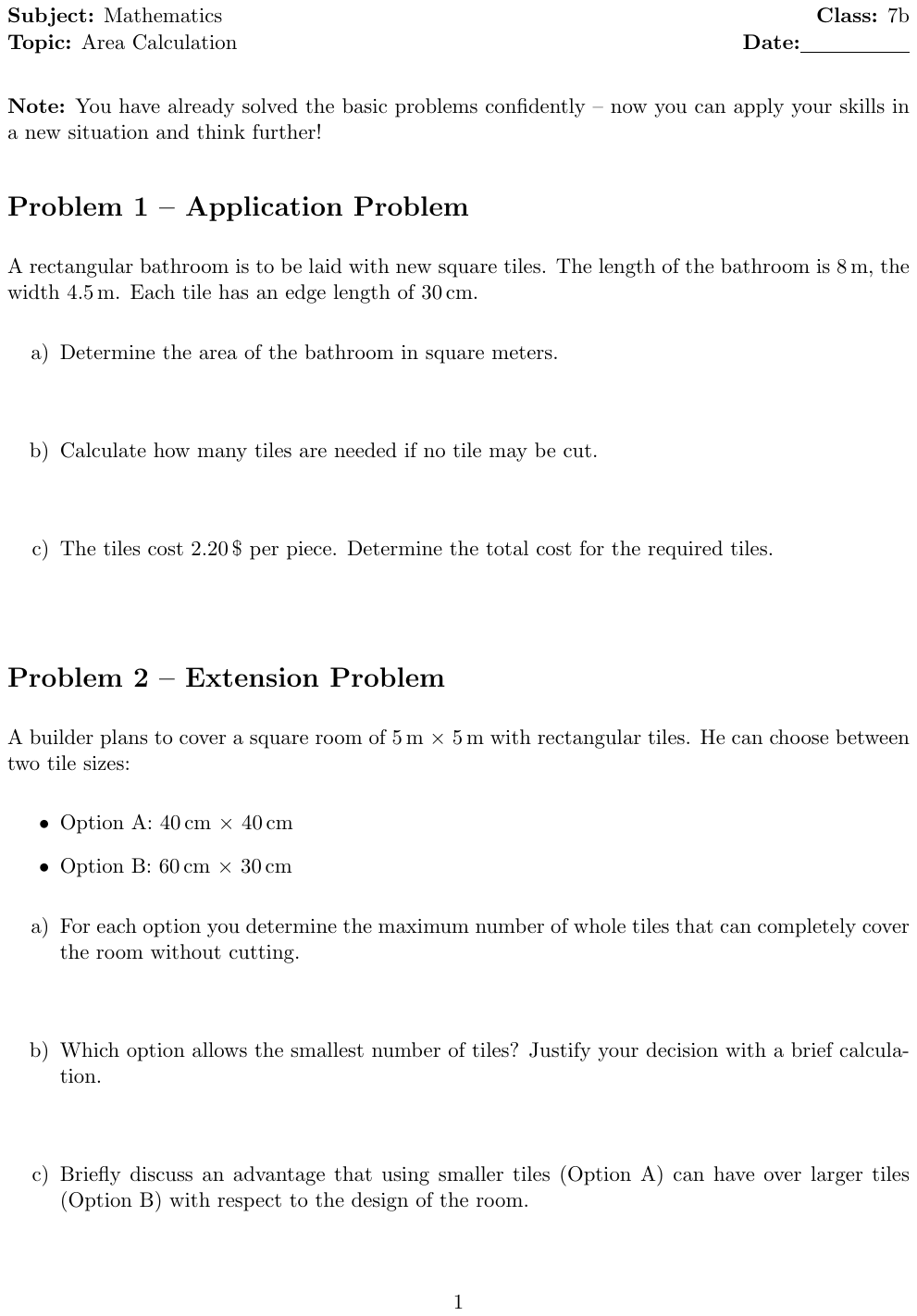}
    \caption{Example worksheet for high performance and low motivation}
    \label{fig:worksheet_example}
\end{figure}

\bibliographystyle{named}
\bibliography{references}

\end{document}